\documentclass[prb,twocolumn,showpacs]{revtex4}
\usepackage{graphicx}
\usepackage{amssymb}
\usepackage{dcolumn}
\usepackage{float}
\usepackage{bm}
\usepackage{multirow}
\usepackage{booktabs}
\usepackage{amsmath} 
\usepackage[T1]{fontenc}
\usepackage[utf8]{inputenc}
\usepackage{lmodern}
\usepackage{makecell}

\newcolumntype{M}[1]{>{\centering\arraybackslash}m{#1}}
\renewcommand{\Re}{\mathrm{Re}\,}

\renewcommand{\Re}{\mathrm{Re}\,}

\newcommand{\bs}{\boldsymbol}
\DeclareMathAlphabet{\bi}{OML}{cmm}{b}{it}
\usepackage{color}

\def\be{\begin{equation}}
\def\ee{\end{equation}}
\def\bearr{\begin{eqnarray}}
\def\eearr{\end{eqnarray}}
\def\la{\langle}
\def\ra{\rangle}

\begin{document}
\title{Optical conductivity of a 2DEG with anisotropic Rashba
interaction at the interface of LaAlO$_3$/SrTiO$_3$}
\bigskip
\author{Alestin Mawrie and Tarun Kanti Ghosh\\
\normalsize
Department of Physics, Indian Institute of Technology-Kanpur,
Kanpur-208 016, India}
\date{\today}
 
\begin{abstract}
We study optical conductivity of a two-dimensional electron gas with anisotropic
$k$-cubic Rashba spin-orbit interaction formed at the LaAlO$_3$/SrTiO$_3$ interface.
The anisotropic spin splitting energy gives rise to different features 
of the optical conductivity in comparison to the isotropic $k$-cubic Rashba spin-orbit 
interaction.
For large carrier density and strong spin-orbit couplings, the density dependence of 
Drude weight deviates from the linear behavior.
The charge and optical conductivities remain isotropic despite anisotropic 
nature of the Fermi contours.
An infinitesimally small photon energy would suffice to initiate inter-band optical
transitions due to degeneracy along certain directions in momentum space. 
The optical conductivity shows a single peak at a given photon energy depending on 
the system parameters and then falls off to zero at higher photon energy. These features 
are lacking for systems with isotropic $k$-cubic Rashba spin-orbit coupling.  
These striking features can be used to extract the information about  
nature of the spin-orbit interaction experimentally and illuminate some light on 
the orbital origin of the two-dimensional electron gas.

\end{abstract}

\maketitle
\section{Introduction}
Spin-orbit interaction \cite{soc,soc2} (SOI) plays an important role in 
understanding physical properties of different materials as it lifts 
the spin degeneracy due to the absence of either the structure inversion 
symmetry or the time reversal symmetry. In general, there are two different 
types of symmetry dependent SOI, Rashba \cite{Rashba1,Rashba2} and Dresselhaus 
\cite{dres} SOIs, in various condensed matter systems. 
In two-dimensional electron gas (2DEG) formed at the III-V semiconductor 
heterostructures\cite{bdas} and in various topological insulating systems \cite{TI}, 
the Rashba SOI (RSOI) is linear in momentum and of the form 
$H_R = i\alpha k_-\sigma_++\textrm{h.c.}$, where $\alpha$ is the strength of RSOI, 
$\sigma_\pm = \sigma_x \pm i \sigma_y$ with $\sigma_x$ and $\sigma_y$ are the Pauli's 
spin matrices and $k_\pm = k_x \pm i k_y$ with $k_x$ and $k_y$ the components of the 
wave vector ${\bf k}$. Besides, the Rashba SOI in two-dimensional hole gas formed at the
interface of p-type GaAs/AlGaAs heterostructures\cite{sherman,winkler}, 
2DEG on the surface of SrTiO$_3$ single crystals \cite{nakamura} and in 2D hole gas formed 
in a strained Ge/SiGe quantum well \cite{strain} is cubic in momentum and is of the 
form $H_R^{\rm iso}=i\alpha k_-^3\sigma_++\textrm{h.c}$. 
The spin splitting energy due to this RSOI is always isotropic 
and hereafter we will mention this as isotropic cubic RSOI.

An extremely high mobility 2DEG was discovered at the interface of the complex oxides 
LaAlO$_3$ and SrTiO$_3$ \cite{nature1,nature2,high-mobility}. The mobility at oxide interfaces is relatively less than that of III-V semiconductor heterostructures\cite{hih-mobilityIIIV}.
Unlike the conventional III-V semiconductor heterojunctions, the
2DEGs at LaAlO$_3$/SrTiO$_3$ interfaces are characterized by
very strong spin-orbit interaction, high carrier densities,
higher effective mass\cite{high-mobility2}. The LAO/STO interface structure now has a broken structure inversion symmetry as a result 
of the confinement along the axis normal to the interface, which leads to the lifting of 
the spin-degeneracy of the six $t_{2g}$ orbitals in STO \cite{lee}. 
Moreover, the $d_{xy}$ orbitals are confined in the $x$-$y$ plane and are localized at 
the interface due to impurities and electron-phonon coupling \cite{popovic}, whereas the electrons associated with the $d_{xz}$ and $d_{yz}$ orbitals \cite{popovic} are itinerant and contribute to transport.
One of the major concerns is to understand the nature of the SOI of the charge carriers at the oxide interface.
In Refs. \cite{held,lee}, a $k$-linear Rashba SOI for $d_{xy}$ orbital and
an isotropic $k$-cubic for $d_{xz}$ and $d_{yz}$ orbitals were proposed.
The magneto-transport measurement of 2DEG at the oxide interface has indicated 
the existence of $k$-cubic RSOI and is modeled using the isotropic $k$-cubic 
SOI, $H_R^{\rm iso}$  \cite{nakamura,nayak}.
On the other hand, the first-principle calculations suggested anisotropic non-parabolic 
spin-split branches for the $d_{xz}$ and $d_{yz}$ orbitals \cite{fasolino}.
Two recent polarization-dependent ARPES revealed non-isotropic Fermi contours 
of the 2DEG at the oxide interface \cite{arpes}.
Very recent theoretical study \cite{xiao} predicted that these orbitals are characterized by 
$k$-cubic but anisotropic Rashba spin-orbit interaction whose form is given by
$ H_{R}^{\rm ani} = \alpha (k_{x}^2 - k_{y}^2)({\bf k} \times {\bs \sigma}) \cdot \hat z $. 
The spin splitting energy and Fermi contours become highly anisotropic as a result of this
anisotropic SOI. In this paper we will refer this as anisotropic RSOI.  
This form of the anisotropic RSOI \cite{xiao} enables to explain the experimental 
observations of the anisotropic spin susceptibility \cite{spin-sus,spin-sus1} successfully. 
It is also shown \cite{xiao} that the anisotropic RSOI leads to different behavior 
of the spin Hall conductivity, in comparison to the isotropic $k$-cubic RSOI.

The spectroscopic measurement of the absorptive part of the optical 
conductivity can probe the spin-split energy levels. 
Theoretical studies of the optical conductivity of various charged systems 
with an isotropic $k$-cubic Rashba SOI have been carried out \cite{lin,yang,wong,li,mawrie}. 
It is shown that the optical transition takes place for 
a certain range of photon energy depending on the carrier density and spin-orbit 
coupling constant. At zero temperature, it takes a box-like function and its value is
$\sigma_{xx}^{\rm iso} = 3e^2/(16 \hbar)$, independent of carrier density and spin-orbit
coupling strength.

In this paper we study the Drude weight and optical conductivities of the 
2DEG with anisotropic $k$-cubic RSOI formed at the oxide interface and compare 
our results with that of the isotropic $k$-cubic RSOI. 
Firstly, we present the characteristics of the zero-frequency Drude weight 
as a function of the charge density and strength of the anisotropic RSOI.
We find that the Drude weight is strongly modified due to the presence of
the anisotropic $k$-cubic SOI. It deviates from the linear density dependence
for large carrier density and for strong spin-orbit coupling. The Drude weight decreases
with the increase of the strength of RSOI. 
Secondly, we find that an infinitesimally small photon energy would initiate 
the inter-band optical transition. This is due to the vanishing spin-splitting 
energy along certain directions in the momentum space. 
There is a single peak in the optical conductivity and its value depends
on the electron density and strength of the anisotropic RSOI.
Moreover, the charge and optical conductivities are isotropic despite the 
fact that the RSOI is anisotropic.
In conventional 2DEG the van Hove singularities largely affects the various physical properties like transport$^{32}$, character of plasmons$^{33}$, N-type kink in photoluminescence$^{34}$, dilute-magnetic semiconductor properties$^{35}$ etc.
Here as well, the van Hove singularities drastically affects the optical conductivity, thereby responsible for the single peak observed in it. The van Hove singularities are of the same $M_1$ type.
These features can be used to find out the nature of the RSOI experimentally.

This paper is organized as follows. In section II, we describe basic properties
of the 2DEG with anisotropic $k$-cubic spin-orbit interaction.
In section III, we present the analytical and numerical results of 
the Drude weight and the optical conductivity. 
The summary and conclusions of this paper are presented in section IV.

\section{Description of the Physical System}
The effective Hamiltonian of the electron in $d_{xz}$ and $d_{yz}$ orbitals 
at the interface of LAO/STO is given by\cite{xiao}
\begin{eqnarray}
H = \frac{\hbar^2 {\bf k}^2}{2m^*} + 
\alpha (k_x^2-k_y^2)({\bf k} \times {\bs \sigma}) \cdot \hat z,
\end{eqnarray} 
where $m^\ast$ is the effective mass of the electron, $\alpha$ is the strength of 
the anisotropic RSOI and ${\bs\sigma}=\sigma_x \hat{x}+\sigma_y\hat{y}$.
The above Hamiltonian is valid within the narrow region around the $\Gamma$ point.
The anisotropic dispersion relations and  the corresponding eigenfunctions are given by
\begin{eqnarray}\label{spec}
E_\lambda({\bf k})=\frac{\hbar^2 k^2}{2m^\ast} + \lambda\alpha k^3 |\cos 2\theta|
\end{eqnarray}
and 
$ \psi_{\bf k}^\lambda({\bf r}) = e^{i {\bf k} \cdot {\bf r}} 
\phi_{\bf k}^\lambda({\bf r})/\sqrt{\Omega}$
with the spinor
\begin{eqnarray}
\phi_{\bf k}^\lambda({\bf r}) = \frac{1}{\sqrt{2}} 
\begin{pmatrix}
1\\
\lambda \eta_{\bf k} i e^{i\theta}
\end{pmatrix}.
\end{eqnarray}
Here $\Omega$ is the surface area of the two-dimensional system, 
$\lambda=\pm$ denotes the spin-split branches and 
$\eta_{\bf k}=\cos 2\theta/|\cos 2\theta|$ 
with $\theta = \tan^{-1}(k_y/k_x)$ measures the anisotropy of the spectrum.
The magnitude of the anisotropic spin-splitting energy is 
$E_g({\bf k}) = | E_+({\bf k}) - E_-({\bf k}) | = 2\alpha k^3|\cos 2\theta|$.
The spin splitting energy vanishes at $\theta= (2p+1)\pi/4$ with
$p=0,1,2,3$. On the other hand, the maximum spin-splitting 
($E_{g}^{\rm max} = 2 \alpha k^3$) occurs at $\theta=p\pi/2$.
To allow only the bound states, the wave-vector ${\bf k}$ should have an upper cut-off given 
by $k_c(\pi/4)=\hbar^2/(3m^*\alpha) $ which corresponds to the cut-off energy 
$ E_c= \alpha k_c^3/2$.

The spin texture on the $k_x$-$k_y$ plane can be obtained from the
average values of spin vector (in units of $3\hbar/2$) 
$ {\bf P}_{\lambda}({\bf  k}) = \la {\bs \sigma} \ra_{\lambda} =
\lambda  \eta_{\bf k} \hat \theta $, where
$\hat \theta = -\hat x \sin \theta + \hat y \cos \theta $ is the unit
polar vector.
The electron spin lies in the ${\bf k} $ plane and always locked at
right angles to its momentum.

The Berry connection \cite{berry} is defined as 
${\bf A_k} = i \langle \phi_{\bf k}^\lambda|\bigtriangledown_{\bf k}|\phi_{\bf k}^\lambda\rangle$,
where $ \phi_{\bf k}^{\lambda} $ is the spinor part of the wave 
function $\psi_{\bf k}^{\lambda}({\bf r})$. The Berry connection for this 
system yields $ {\bf A}_{\bf k} = -\hat \theta/(2k)$.
Using the expression of the Berry phase \cite{berry} $\gamma = \oint {\bf A_k} \cdot d{\bf k} $, 
we get $ \gamma_{\rm ani} = - \pi $ for anisotropic case, 
whereas $\gamma_{\rm iso} = 3 \pi $ for isotropic cubic RSOI \cite{mawrie}.

\begin{figure}[!htbp]
\begin{center}\leavevmode
\includegraphics[width=95mm,height=70mm]{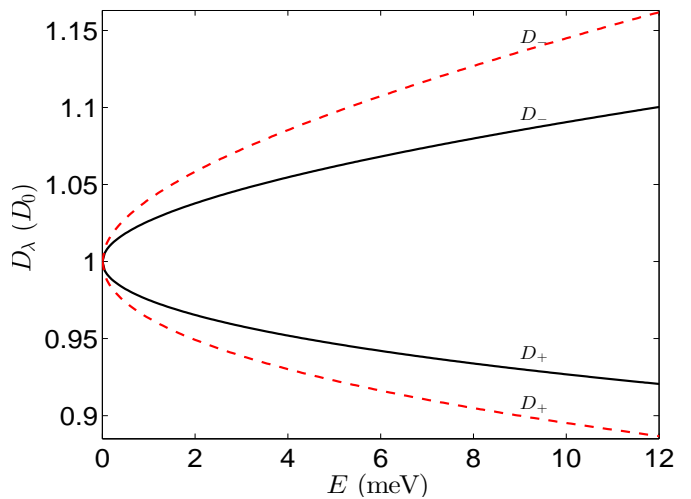}
\caption{(color online) Plots showing the density of states in the units of 
$D_0$ for two different values of $\alpha $. 
Here $\alpha=0.004$ eV nm$^3$ (solid black) and $\alpha=0.006$ eV nm$^3$ (dotted red).}
\label{Fig1}
\end{center}
\end{figure}

In order to obtain two anisotropic Fermi contours $k_{f}^{\lambda}(\theta)$,
we need to calculate density of states (DOS) and Fermi energy $E_f$.
The density of states of the spin-split energy branches are given by
\begin{eqnarray}
D_\lambda(E) & = &
\int \frac{d^2k}{(2\pi)^2} \delta(E - E_{\lambda}({\bf k})) \nonumber \\
& = & \frac{D_0}{2\pi}\int_0^{2\pi} 
\frac{ k_E^\lambda(\theta)\,d\theta}{|k_E^\lambda(\theta) + 
\lambda 6 \pi \alpha D_0 {(k_E^\lambda(\theta))}^2 |\cos 2\theta||} \nonumber,
\end{eqnarray}
where $D_0=2\pi m^\ast/h^2$
and $k_E^\lambda(\theta)$ being the solution of the equation
$( \hbar k_E^\lambda)^2/2m^\ast +\alpha \lambda (k_E^\lambda)^3 |\cos 2\theta| - E = 0$.
The density of states is obtained numerically and their
characteristics for the two branches are shown in Fig. 1. 
The DOS of the anisotropic spin-split levels varies asymmetrically
with respect to $D_0$.
For fixed electron density $n_e$ and $\alpha$, 
the Fermi energy ($E_f$)  
is obtained from the conservation of electron number 
$n_e = \int_{0}^{E_f} \sum_\lambda D_\lambda(E) dE$.
The variations of the Fermi energy with $n_e$ and $\alpha$ are shown in
Fig. 2.
The Fermi energy increases with the increase of the carrier density.
On the other hand, the Fermi energy decreases with the increase of the
spin-orbit coupling strength. 
The Fermi wave vectors $k_{f}^{\lambda}(\theta) $ can be obtained numerically 
from the solutions of the equation $\hbar^2 k^2/2m^\ast + \lambda \alpha
k^3 |\cos 2\theta| - E_f = 0$.
The Fermi contours are depicted in Fig. 5 (color: black). 
\begin{figure}[!htbp]
\begin{center}\leavevmode
\includegraphics[width=90mm,height=65mm]{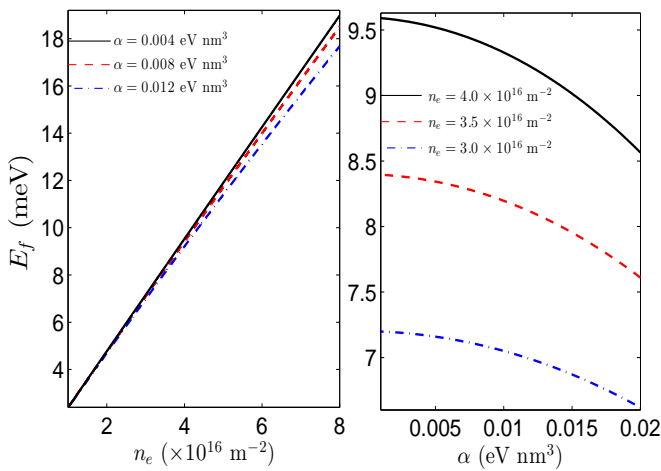}
\caption{(color online) Plots of the Fermi energy vs $n_e$ and $\alpha$.
Left panel: plots of the Fermi energy vs density 
for different values of $\alpha$. 
Right panel: plots of the Fermi energy 
vs $\alpha $ for different values of carrier density $n_e$.}
\label{Fig1}
\end{center}
\end{figure}

\section{Drude weight and Optical conductivity}
The complex charge conductivity for a two-level system of charge 
carriers in presence of a sinusoidal electric field 
(${\bf E}(\omega) \sim \hat {\bf x} E_0 e^{i\omega t} $) can be written as 
$
\Sigma_{xx}(\omega) = \sigma_D(\omega) + \sigma_{xx}(\omega),
$
where $ \sigma_D(\omega)$ is the intra-band induced dynamic Drude conductivity and 
$\sigma_{xx}(\omega) $ is the inter-band induced complex optical conductivity.

The absorptive part of the conductivity can be obtained by taking the real part 
of $\Sigma_{xx}(\omega)$ and is given by
$$\Re [\Sigma_{xx}(\omega) ] = D_w\delta(\omega) + \Re [\sigma_{xx}(\omega)].$$
Here, $D_w$ is known as the Drude weight measuring the 
Drude conductivity ($\sigma_d = \tau D_w/\pi$) for a DC electric field 
and $ \Re [\sigma_{xx}(\omega)] $ is the optical conductivity 
as a function of the frequency of the AC electric field with vanishing momentum 
$q\rightarrow 0$. The vanishing momentum of the electric field forces  
the charge carriers to make a transition from $\lambda=-1$ branch to $\lambda = +1$ branch 
such that the momentum is conserved.

{\bf Drude weight}: The semi-classical expression for the Drude weight 
at low temperature is given by \cite{mahan}
\begin{eqnarray}
D_w = \pi e^2 \sum_\lambda \int \frac{d^2{\bf k}}{(2\pi)^2} \, 
\langle \hat{v}_x\rangle_{\lambda}^2 \delta(E_\lambda({\bf k})-E_f).
\end{eqnarray}
Here $ \hat{v}_x $ is the $x$-component of the velocity operator.
Using the Heisenberg's equation of motion, $i \hbar \dot {\bf r} = [{\bf r}, H] $,
the $x$- and $y$-components of the velocity operator are given by
\begin{eqnarray} \label{vx}
\hat{v}_x & = &
\frac{\hbar k_x}{m^\ast} \mathbb{I} +
\frac{\alpha}{\hbar}\big[(3k_x^2-k_y^2)\sigma_y-2k_xk_y\sigma_x\big]
\end{eqnarray}
and
\begin{eqnarray}
\hat{v}_y &=&\frac{\hbar k_y}{m^\ast} \mathbb{I} - \frac{\alpha}{\hbar}
\big[(3k_y^2-k_x^2)\sigma_x-2k_xk_y\sigma_y\big].
\end{eqnarray}

For the system with anisotropic cubic RSOI, the calculation of the Drude
conductivity yields
\begin{eqnarray}
D_{w}^{\rm ani} & = & \Big(\frac{e}{2\pi \hbar }\Big)^2 
\sum_\lambda \int_{0}^{2\pi} 
m^*[v_f^\lambda(\theta)]^2 B_{\lambda}(\theta) d\theta,
\end{eqnarray}
where $ v_f^{\lambda}(\theta) = \hbar k_f^{\lambda}(\theta)/m^* $ and
\begin{eqnarray}  
B_{\lambda}(\theta) = 
\frac{[\cos\theta + \lambda \eta_{\bf k}\alpha V_{f}^{\lambda}(\theta) 
(5\cos\theta +  \cos 3\theta)/2]^2}{1 + \lambda 3 \alpha V_f^{\lambda}(\theta)}
\end{eqnarray}
with $V_f^{\lambda}(\theta) = (m^*/\hbar^2) k_f^{\lambda}(\theta) $.

\begin{figure}[!htbp]
\begin{center}\leavevmode
\includegraphics[width=102mm,height=75mm]{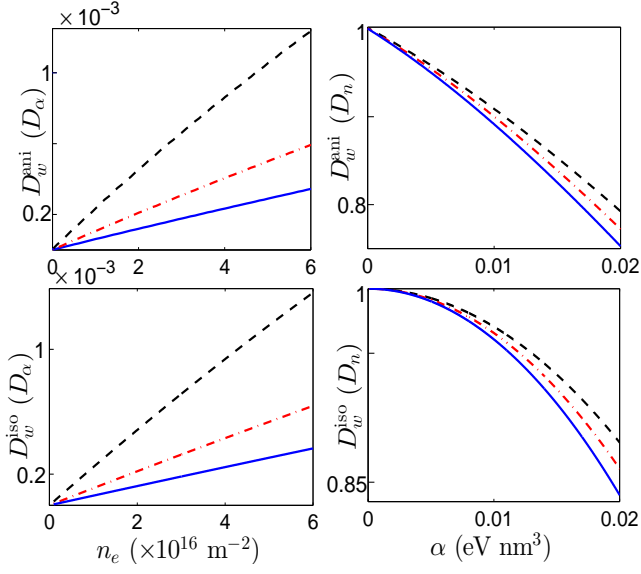}
\caption{(color online) Left panels: plots of the Drude weight $D_w$ 
(in units of $D_{\alpha} = \pi e^2/m^* l_{\alpha}^2 $) vs 
$n_e$ for $\alpha = 0.006$ eV nm${}^3$ (solid: blue),
$\alpha=0.008$ eV  nm${}^3$ (dotted-dashed: red)
and $\alpha=0.012$ eV nm${}^3$ (dashed: black). Here
$l_{\alpha} = m^* \alpha/\hbar^2$.
Right panel: plots of $D_w$ (in units of $D_n = \pi e^2 n_e/m^*$)
vs $\alpha $ for $n_e = 3\times 10^{16}$ m$^{-2}$ (dashed: black),
$n_e=3.5\times 10^{16}$ m$^{-2}$ (doted-dashed: red) and
$n_e=4 \times 10^{16}$ m$^{-2}$ (solid: blue).}
\label{Fig2}
\end{center}
\end{figure}

For carrying out the numerical calculation,
we adopt the following parameters used in Refs. \cite{held,xiao}: 
$n_e=3.5 \times 10^{16}$ m$^{-2}$ and $m^\ast/m_0=1$, where $m_0$ is the
bare mass of the electron.
In Fig. 3, the variations of the Drude weight with the carrier density and 
with the strength of the Rashba spin-orbit interactions are shown. 
The plots of the Drude weight vs carrier density for three different values
of $\alpha $ are shown in the left panels of Fig. 3.                    
The analytical expression of $D_{w}^{\rm iso}$ obtained in Ref. 
\cite{mawrie} 
clearly shows the deviation from the linear density dependence.
Because of the small value of $\alpha$ considered here, the deviation is not visible
in this figure.
On the other hand, the Drude weight vs $\alpha $ for three different values 
of carrier density are plotted in the right panels of Fig. 3. 
The Drude weight decreases with the increase of
$\alpha$ but the decreasing nature of $D_w$ for the two different cases is
quite different. This important feature would help to know the nature of the RSOI.

{\bf Optical Conductivity}:
The generalized Kubo formula of the optical conductivity in terms of 
the Matsubara Green's function is given by\cite{mahan}
\begin{eqnarray}\label{NMac}
& &\sigma_{\mu \nu}(\omega)=  -\frac{e^2 T}{i\omega}\frac{1}{(2\pi)^2} 
\int  d^2 {\bf k} \nonumber\\
& \times &  \sum_l {\rm Tr}\langle  \hat v_\mu \hat{G}({\bf k},\omega_l) \hat v_\nu 
\hat{G}({\bf k},\omega_s+\omega_l) \rangle_{i\omega_s\rightarrow \omega + i\delta }.
\end{eqnarray}
Here, $\mu, \nu =x,y$, $T$ being the temperature, $\omega_s=(2s+1)\pi T$ and $\omega_l=2l\pi T$ 
are the fermionic and bosonic Matsubara frequencies with $s$ and $l$ are integers, respectively.

The matrix Green's function associated with the Hamiltonian 
given by Eq. (1) is 
\begin{eqnarray} \label{gf}
G({\bf k},\omega_n) = \frac{1}{2}\sum_\lambda 
\Big[ \mathbb{I} +  {\bf P}_{\lambda}({\bf k}) 
\cdot {\bs \sigma}\Big]G_0^{\lambda}({\bf k},\omega_n).
\end{eqnarray}
Here $\mathbb{I}$ is a $2 \times 2$ unit matrix 
and $G_0^{\lambda}({\bf k},\omega_n)=1/(i\hbar\omega_n + \mu_0 - E_{\lambda}({\bf k}) ) $ 
with $\mu_0$ being the chemical potential.
It indicates that the optical spectral weight is directly related to the
local spin texture ${\bf P}_{\lambda}({\bf k})$.

Substituting Eqs. (\ref{vx}) and (\ref{gf}) into Eq. (\ref{NMac}),
the $xx$-component of the longitudinal conductivity reduces to 
\begin{eqnarray} \label{op-con}
\sigma_{xx}(\omega) & = -& \frac{e^2}{i(2\pi \hbar)^2\omega}
\int_{0}^{\infty} \int_{0}^{2\pi} \alpha^2 k^5 \, \cos^22\theta\sin^2\theta dk d\theta \nonumber\\
& \times &\Big[\frac{f(E_-)-f(E_+)}{\hbar\omega + i\delta - E_+ + E_-} +
(E_-\leftrightarrow E_+)\Big],
\end{eqnarray}
where $f(E)=[e^{(E-\mu_0)\beta} + 1]^{-1}$ is the Fermi-Dirac distribution function
with $\beta = 1/(k_BT) $.

We have carried out the same calculation for other components of the conductivity
tensor $\sigma_{\mu\nu}(\omega)$.
We find that $ \sigma_{yy}(\omega) = \sigma_{xx}(\omega) $ and
$\sigma_{xy}(\omega) = \sigma_{yx}(\omega) = 0 $. Hence the optical 
conductivity remains isotropic despite the fact that the Fermi contours 
are anisotropic.

Using the fact that $\omega >0$ and after performing the $k$ integral, 
the expression for the absorptive part of the optical conductivity 
at $T=0$ is given by
\begin{eqnarray}\label{cond}
\Re [\sigma_{xx}(\omega) ] =\frac{e^2}{24 h} 
\int_0^{2\pi}d\theta \sin^2\theta \Big[ \Theta(\mu_{+}) - \Theta(\mu_-) \Big],
\end{eqnarray}
where $\Theta(x) $ is the unit step function and $ \mu_{\pm} = E_{\pm}(k_{\omega}) - \mu_0 $ 
with $k_\omega \equiv k_w(\theta) = (\hbar\omega/2\alpha|\cos 2\theta|)^{1/3}$.
This integral cannot be solved analytically due to $\theta$ dependence
of $k_{\omega}$.

\begin{figure}[!htbp]
\begin{center}\leavevmode
\includegraphics[width=150mm,height=125mm]{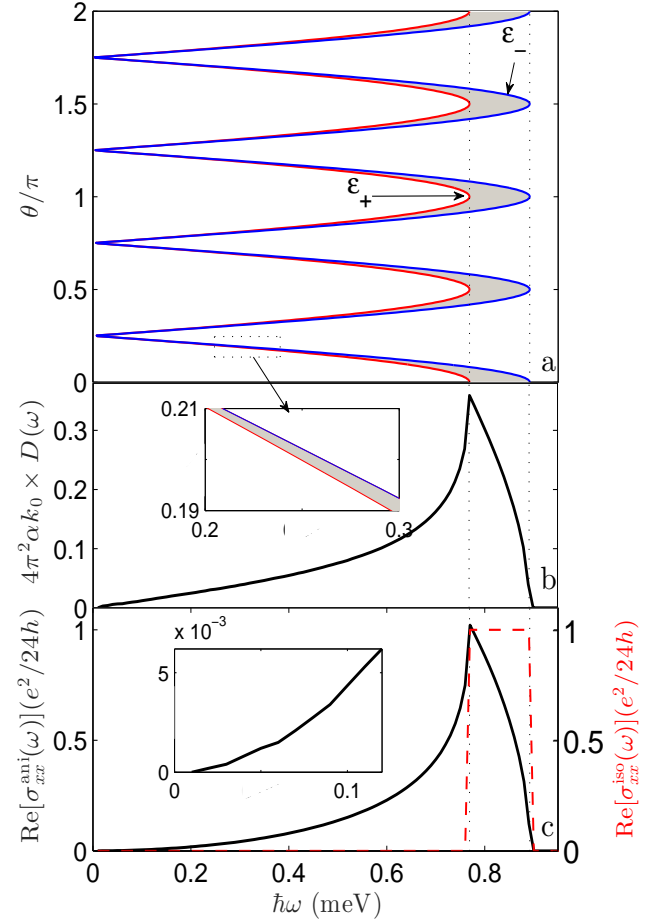}
\caption{(color online): Top panel: Plots of $\epsilon_{\pm}(\theta) $ 
vs $\theta $.
Middle panel: plots of the joint density of states
vs $\hbar\omega$ with $k_0=\sqrt{2\pi n_e}$. 
Bottom panel: the real part of the optical conductivity as a function of photon 
energy $\hbar\omega $.}
\label{Fig2}
\end{center}
\end{figure}

On the other hand, in isotropic cubic Rashba SOI the closed form expression of 
the absorptive part of the optical conductivity at $T=0$ K is given by\cite{mawrie}
\begin{eqnarray}\label{isotropic}
\Re [\sigma_{xx}^{\rm iso}(\omega)]=\frac{3e^2}{16 \hbar}
\big[ \Theta(\tilde \mu_+) - \Theta(\tilde \mu_-) \big],
\end{eqnarray}
where $ \tilde \mu_{\pm} = E_{\pm}(\tilde k_{\omega}) - \mu_0 $
with $\tilde k_\omega=(\hbar\omega/2\alpha)^{1/3} 
= k_w(\theta = (2p+1)\pi/2)$.
It leads to featureless optical conductivity which has box shape with 
the height
$ \sigma_{xx}^{\rm iso} = 3e^2/(16 \hbar)$ which is independent
of the carrier density and $\alpha$. 
Note that simultaneous presence of isotropic Rashba and Dresselhaus 
SOI leads to anisotropic Fermi contours, in turns produces interesting optical
features. Whereas anisotropic RSOI alone gives rise to anisotropic 
Fermi contours and provides distinct optical features.
\begin{figure}[!htbp]
\begin{center}\leavevmode
\includegraphics[width=80mm,height=70mm]{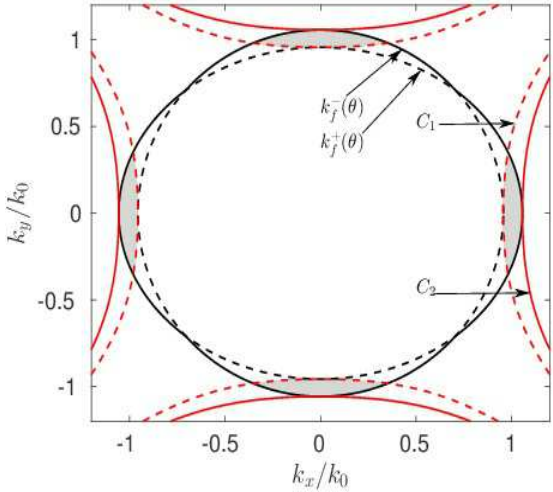}
\caption{(color online)
Plots of the Fermi contours $k_f^+(\theta) $, $k_f^-(\theta) $,
the constant-energy difference curves $C_1$: $ E_g({\bf k}) = \epsilon_1 $
and $C_2$: $ E_g({\bf k}) = \epsilon_2 $.}
\label{Fig1}
\end{center}
\end{figure}

Here we shall present how the anisotropic RSOI alone gives rise 
to some unique features of the optical conductivity. 
We first evaluate $ \Re [\sigma_{xx}(\omega)] $ numerically 
using the parameters $\alpha = 0.004$ eV nm$^3$, 
$n_e = 3.5 \times 10^{16}$ m$^{-2}$ and $m^\ast/m_0=1$  as used in Refs. \cite{held,xiao} 
and shown in the lower panel of Fig. 4. 
For comparison with the isotropic case, we plot $ \Re [\sigma_{xx}^{\rm iso}(\omega)] $ 
which appears as the rectangular box on the right side of the lower panel of Fig. 4.
We depict $ \epsilon_{\pm}(\theta) = 2\alpha [k_{f}^{\pm}(\theta)]^3 |\cos(2\theta)| $ in
the top panel of Fig. 4. The contribution to optical conductivity arises from the shaded
angular region.
The optical transitions from $\lambda = - 1$ to
$\lambda = + 1 $ occur when the photon energy satisfies the inequality
$ 0 < \hbar \omega < \epsilon_{-}(\theta)$.
One can see that an infinitesimally small photon energy can initiate 
the optical transition, in 
complete contrast to the isotropic SOI case. This is due to the presence of 
the degenerate lines $ \theta = (2 p+ 1)\pi/4 $.
There is a single peak of the Re $[\sigma_{xx}(\omega)]$ at 
$\hbar\omega = \epsilon_+(p\pi/2) = 2\alpha {[k_{f}^{+}(p\pi/2)]}^3 $ and the optical
conductivity becomes zero when 
$\hbar\omega \geq \epsilon_{-}(p\pi/2) =  2\alpha{[k_{f}^{-}(p\pi/2)]}^3 $.
For better understanding of these features, we plot the constant energy-difference 
curves $E_g({\bf k})  = \epsilon_{\omega} $ for 
$ \epsilon_{\omega} = \epsilon_{+}(p \pi/2)=\epsilon_1 $  ($C_1$: dashed) 
and  $\epsilon_{\omega} = \epsilon_{-}(p\pi/2)=\epsilon_2$  ($C_2$: solid) in Fig. 5.
The area intercept by the curves $C_i$ with $i=1,2$ and the Fermi contours 
($k_{f}^{\lambda}$) are responsible for the ${\bf k}$-selective 
optical transitions as shown in Fig. 5.
It should be noted that the anisotropic $k$-cubic band is well separated from the other two bands with $k$-linear SOI for the parameters used in Refs.\cite{held,xiao}. As a result, the contribution to the optical conductivity from other two bands having $k$- linear SOI is ruled out since they occur at $\hbar\omega$ much larger than $\epsilon_-(p\pi/2)\approx 0.9$ meV.

The overall behavior of the optical spectra can be understood 
from the joint density of states which is 
given as
\begin{eqnarray*}
D(\omega)= \int \frac{ d^2 {\bf k}}{(2\pi)^2} 
[f(E_+({\bf k}))-f(E_-({\bf k}))] \delta(E_g({\bf k}) - \hbar\omega).
\end{eqnarray*}

\begin{figure}[!htbp]
\begin{center}\leavevmode
\includegraphics[width=95mm,height=70mm]{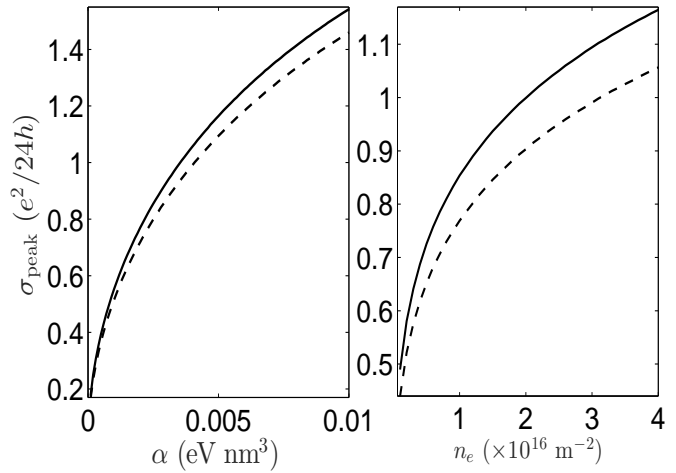}
\caption{(color online) Left panel:
Plots of $\sigma_{\rm peak} $ vs $ \alpha $ for fixed values of
$n_e = 4.0 \times 10^{16} $ m$^{-2}$ (dashed) and
$n_e = 3.0 \times 10^{16} $ m$^{-2}$ (solid).
Right panel: Plots of $\sigma_{\rm peak} $ vs $n_e$ for different values of
$\alpha = 0.004$ eV nm$^3 $ (solid) and
$\alpha = 0.006$ eV nm$^3 $ (dashed). }
\label{Fig2}
\end{center}
\end{figure}

\begin{figure}[!htbp]
\begin{center}\leavevmode
\includegraphics[width=90mm,height=70mm]{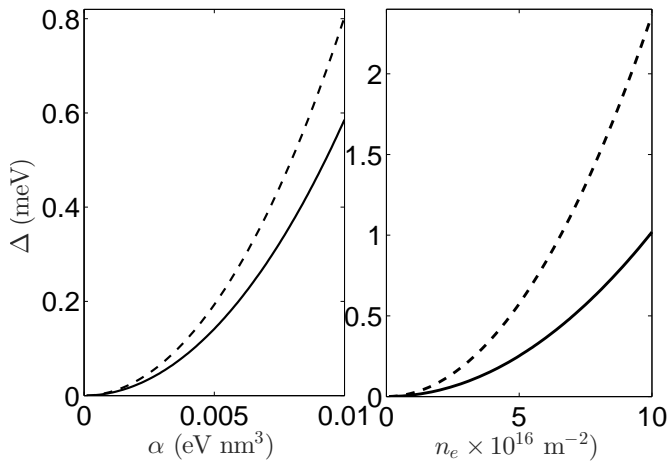}
\caption{(color online) Left panel:
Plots of $\Delta $ vs $ \alpha $ for fixed values of
$n_e = 3.5 \times 10^{16} $ m$^{-2}$ (dashed) and
$n_e = 3.0 \times 10^{16} $ m$^{-2}$ (solid).
Right panel: Plots of $\Delta $ vs $n_e$ for different values of
$\alpha = 0.004$ eV nm$^3 $ (solid) and
$\alpha = 0.006$ eV nm$^3 $ (dashed). }
\label{Fig2}
\end{center}
\end{figure}

It can be reformulated as 
\begin{eqnarray} \label{jdos}
D(\omega)=\frac{1}{(2\pi)^2}\int_C \frac{dC[f(E_+(k_\omega))-f(E_-(k_\omega))]}
{|\partial_k E_g({\bf k})|_{E_g=\hbar\omega}}.
\end{eqnarray}
Here $C$ is the line element along the contour. 
The joint density of states vs $\hbar\omega$ is plotted in the middle panel of Fig. 4. 
The location of the single peak and the region of zero optical conductivity are
nicely described by the joint density of states. 
It can be seen from Eq. (\ref{jdos}) that any peak may arise
whenever $|\partial_k E_g({\bf k})| $ attains a minimum value.
For the present problem, the singular points are at  ${\bf k}_s = (k, p\pi/4)$. 
The single peak appears at $\epsilon_{\omega} = \epsilon_{+}(p \pi/2)$ in the joint 
density of states corresponds to the well known van Hove singularity.
The asymmetric spin-splitting at the Fermi contours along the $ k_y = k_x = 0 $ lines is
the reason for the appearance of the peak at $ \epsilon_{+}(p\pi/2) $.

There are three different types of the singularity  \cite{ssp-book} depending on the nature
of change of the energy gap around the singular points ${\bf k}_s $.
Using the Taylor series expansion of  $ E_g({\bf k})$ around ${\bf k}_s$ as
$ E_g({\bf k}) = E_g({\bf k}_s) + \sum_{\mu} a_\mu(p) (k_\mu - k_{s\mu})^2 $ 
with the expansion coefficients 
$ 2a_\mu(p) = \frac{\partial^2 E_g({\bf k})}{\partial k_{\mu}^2}\Big|_{{\bf k}_s} $.
The co-efficients $a_\mu$ are as follows
$ a_x(p) = \alpha k [5 + 7(-1)^p] $ and $ a_y(p) = \alpha k [5 - 7 (-1)^p]$.
The sign of the coefficients will determine the type of classification of 
the various singular points. 
One can easily find that the signs of $a_{x}$ and $a_y$ at different singular points 
are $(-1)^p $ and $(-1)^{p+1}$, respectively.
Therefore, every singularities are all of the same class
i.e. $M_1$ type.

The variations of the peak height $(\sigma_{\rm peak})$ with $n_e$ and 
$\alpha $ are shown in Fig. 6.
It strongly depends on the Fermi energy.
We also define a width $\Delta = \epsilon_-(p\pi/2) - \epsilon_+(p\pi/2)$, the
difference between peak position and the position beyond which $\sigma_{xx}(\omega) $ 
vanishes. Its variation with $\alpha$ as well as $n_e$ are shown in Fig. 7. 
It shows that $\Delta $ increases with the increase of $n_e$ as well as $\alpha  $.

\section{Summary and Conclusion}
We have studied the Drude weight and optical conductivity for 
2DEG with $k$-cubic anisotropic RSOI at the oxide interface.
We have presented the variation of the zero-frequency Drude weight with
the carrier density as well as  the strength of the anisotropic spin-orbit
coupling. 
For anisotropic RSOI, the Drude weight deviates from the linear density
dependence.
It is indicated that the spectral weight is directly related to the local
spin texture in momentum space.
We found that the charge and optical conductivities remain isotropic 
although the Fermi contours are anisotropic.
It is found that an infinitesimally small photon energy can trigger inter-band 
optical conductivity. This is due to the fact that the spin-splitting energy vanishes
along the certain directions in ${\bf k}$ space. 
We found a single peak in the optical conductivity whose value depends
on the Fermi energy.
We have shown that the van Hove singularities responsible for the single 
peak in the optical conductivity are of the same $M_1$ type.
The different features of the conductivity can determine the information 
of the nature of the spin-orbit interaction experimentally and would 
help in understanding the orbital origin of the two-dimensional electron gas
at the oxide interface.


\end{document}